\documentstyle{mn}                 %
\def\kms{km\,s$^{-1}$}

\def\Ha{H$\alpha$}

\def\lam{$\lambda$}

\input{psfig.tex}

\title[Supernova 1991D]
{The Exceptionally Bright Type Ib Supernova 1991D.
\thanks{Based on observations collected at the European Southern Observatory,
Chile, ESO4-004-45K}}

\author[Benetti et al.]
{S. Benetti$^{1}$, D. Branch$^{2}$, M. Turatto$^{1}$,
E. Cappellaro$^{3}$, E.~Baron$^{2}$, L. Zampieri$^{1}$, \and M. Della
Valle$^{4}$, A. Pastorello$^{1,2}$
\\
$^1$INAF - Osservatorio Astronomico di Padova, vicolo dell'Osservatorio 5,
I-35122 Padova, Italy \\
$^2$Department of Physics and Astronomy, University of Oklahoma,
Norman, OK 73019, USA\\
$^3$INAF - Osservatorio Astronomico di Capodimonte, via Moiariello 16,
I-80131 Napoli, Italy\\
$^4$INAF - Osservatorio Astrofisico di Arcetri, Largo E. Fermi 5,
I-50125 Firenze, Italy\\
}

\date{Received ................; accepted ................}

\begin{document}

\maketitle

\begin{abstract}

Photometric and spectroscopic observations of the peculiar Type~Ib
Supernova~1991D are presented. SN~1991D was exceptionally bright for a
Type~Ib supernova. The He~I lines were rather weak and the velocity at
the photosphere as a function of time was unusually low.  Comparison
of the observed and synthetic spectra indicates that either hydrogen
was ejected with a minimum velocity of 12,000 \kms or the spectrum
contained features caused by lines of Ne~I. Light curve modelling
suggests that the progenitor probably had a very large radius ($\sim
10^{14}$ cm) and that a considerable amount of $^{56}$Ni was
synthesized during the explosion ($\sim 0.7 M_\odot$). We suggest a
progenitor model of SN~1991D that involves a binary system.

\end{abstract} 
 
\begin{keywords} Supernovae: general -- Supernovae: 1991D
\end{keywords}

\section{Introduction} \label{int}

Core collapse in massive stars is believed to be responsible for
supernovae of Types~II, IIb, Ib and Ic. Those of Type~II (SNe~II) are
characterized by the presence of conspicuous hydrogen lines in their
optical spectra. SNe~IIb show strong hydrogen lines around the time of
their maximum brightness, but later these lines become weak or even
disappear, while He~I lines develop.  SNe~Ib develop strong He~I lines
after maximum, while hydrogen lines are weak or absent. SNe~Ic lack
conspicuous hydrogen and He~I lines (as well as the distinctive strong
absorption of SNe~Ia due to Si~II $\lambda\lambda$ 6347, 6371).  For
more details on the supernova spectral classification we refer the
reader to a recent review by Turatto \shortcite{macio}.

\begin{figure}
\psfig{figure=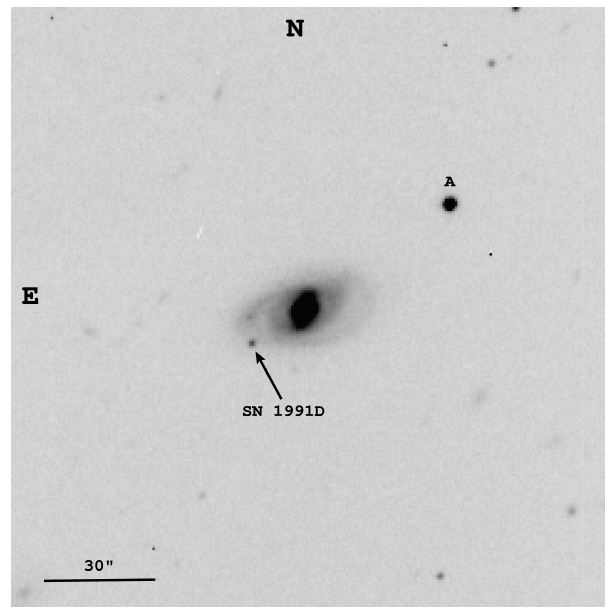,width=8cm,height=8cm}
\caption{SN~1991D in PGC 84044. The image is an R frame taken at
ESO-MPI2.2m telescope on Apr. 17, 1991. The seeing was 1.5 arcsec}
\label{sn}
\end{figure}

Two recent papers (Matheson et al. 2001, hereafter M01; Branch et al
2002, hereafter B02) have advanced our understanding of SN~Ib and Ic
spectra.  M01 presented a large number of optical spectra of SNe~Ib
and Ic, and were able to clarify some of the spectroscopic properties
of these two types.  B02 compared photospheric--phase spectra of 11
SNe~Ib (mostly from M01) with synthetic spectra generated with the
parameterized supernova synthetic-spectrum code SYNOW. They found a
tight relation between the velocity at the photosphere, as determined
by Fe~II features, and the time relative to maximum light.  They also
found the minimum detectable velocity of He~I lines to be usually
higher than 7000 \kms.  Hydrogen appeared to be generally present in
SNe~Ib, with a minimum velocity around 12,000
\kms.

The sample of B02 was restricted to events that showed deep, obvious
He~I absorptions.  B02 mentioned that SN~1991D, an event that had
weaker He~I lines, would be discussed in a separate paper.  In this
paper we present and discuss photometric and spectroscopic
observations that show that this object was an over-luminous and
spectroscopically peculiar SN~Ib that did not conform to the behavior
of the events in the B02 sample.  We also present some comparisons
with SYNOW synthetic spectra that raise the issues of whether lines of
Ne~I are present in SN~1991D, and whether in some SNe~Ib of the B02
sample the strongest optical line of Ne~I, $\lambda 6402$, may be an
alternative to an H$\alpha$ identification.
In Sect. \ref{res} light curve models to V data and a qualitative
progenitor scenario able to explain the main characteristics of SN~1991D
are discussed.

\section{Observations} \label{obs}

SN~1991D was discovered by Remillard et al. \shortcite{remi} on CCD
images obtained on Feb.~6 and 7 UT on a faint spiral arm of PGC~84044,
a Seyfert 1 galaxy.  

Kirshner \shortcite{bob} and Filippenko \& Shields \shortcite{filippo}
classified the event as Type~Ib/c and Ib, respectively, on the basis
of spectra taken on Feb. 10-11 and Feb. 23.  Kirshner also noted that
given the magnitudes reported by Remillard, SN 1991D could be as
luminous as a Type-Ia supernova (see Sect. \ref{phot}).

CCD photometric and spectroscopic observations were obtained at La
Silla on 10 different nights using five different telescopes. Data
reduction followed the standard procedures making use of a PSF fitting
technique for the SN photometric measurements.

\subsection{Photometry} \label{phot}

The measured magnitudes of SN~1991D and the estimated internal errors
(in brackets) are reported in Tab.~\ref{obs_tab}. The internal
consistency of the SN magnitudes in different nights has been checked
mostly using star A of Fig. \ref{sn} (V$=16.41\pm 0.04$,
B$-$V$=0.79\pm0.04$, V$-$R$=0.48\pm 0.06$).\\

\begin{figure}
\psfig{figure=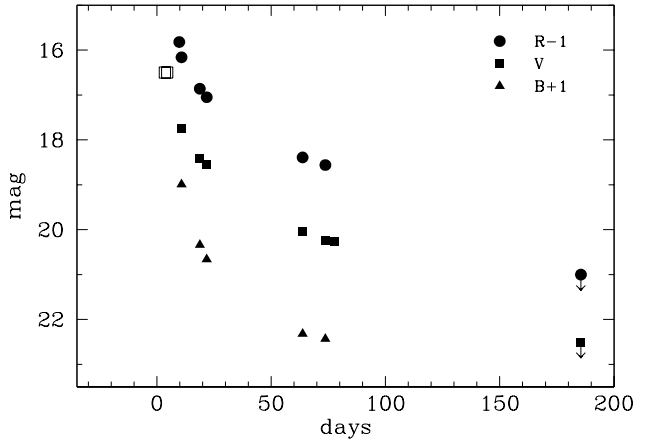,width=9cm,angle=270}
\caption{B, V and R light curves of SN~1991D. The open squares are the
two discovery V magnitudes reported by Remillard et al. (1991). The
phase is counted from the date of V maximum here assumed to have occurred on
Feb. 2, 1991 (see Sect. \ref{phot}).}
\label{phot_fig}
\end{figure}

\begin{figure}
\psfig{figure=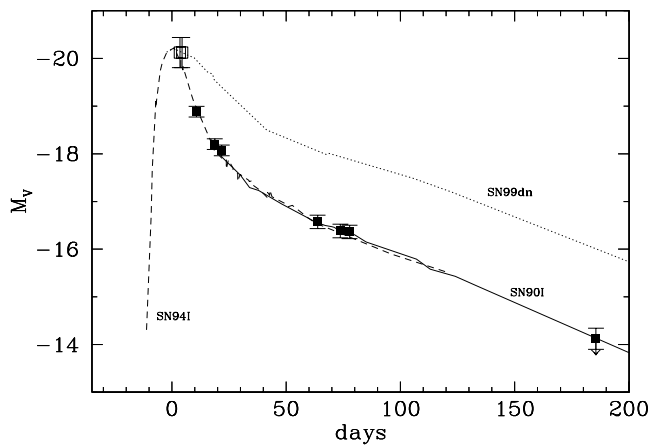,width=9cm,angle=270}
\caption{M$_V$ light curve of SN~1991D compared with those of the
Type~Ib SN~1990I (+0.2 mag) and 1999dn (+3.0 mag, Benetti et al 2002b, in
preparation), and with that of the Type Ic SN~1994I (+2.1 mag)
(Richmond et al. 1996). The absorption (A$_V$) used are: 1.55 mag for
SN 1994I (Ho \& Filippenko, 1995); 0.43 mag and 0.40 mag for SNe~1999dn
and 1990I respectively (Benetti et al 2002b, in preparation). $H_0$
of 65 km s$^{-1}$ Mpc$^{-1}$ is assumed.}\label{abs}
\end{figure}

The B, V and R light curves of SN~1991D are shown in
Fig.~\ref{phot_fig}. The light curves are very fast and show that the
SN was discovered near maximum. The post-maximum (up to +30d) decline
rates are $\beta^{\rm B}_{100}=15.5$\,mag\, (100d)$^{-1}$, $\beta^{\rm
V}_{100}=11.9$\,mag\, (100d)$^{-1}$ and $\beta^{\rm
R}_{100}=9.6$\,mag\, (100d)$^{-1}$, very similar to the well observed
Type~Ic SN~1994I (14.4, 12.0 and $10.8$\,mag\, (100d)$^{-1}$ for the
B, V and R bands respectively; see also
Fig.\ref{abs}).  The decline rates computed on the radioactive tails
using the few points available after phase +50d are:
${\gamma}_{\rm V}\ge2.0$ \,mag~(100d)$^{-1}$ and ${\gamma}_{\rm R}\ge 2.2$
\,mag~(100d)$^{-1}$. During this phase, the optical light curves are
powered by the thermalization of $\gamma$-rays and positrons from the
radioactive decay of $^{56}$Co into $^{56}$Fe. Since in SN~1991D the
decline rates are steeper than the $^{56}$Co decay rate (0.98
mag~(100d)$^{-1}$), the $\gamma$-rays increasingly escaped from the
ejecta without being thermalized, consistent with a low--mass of
the ejecta.

\begin{table*}
\caption{Photometric measurements of SN~1991D}\label{obs_tab}
\begin{flushleft}
\begin{tabular}{lcccccl}
\hline
    date & J.D.     &  B     &    V         &    R         & instr.\\
         & 2400000+ &        &              &              &       \\
\hline
12/2/91& 48299.80 &          &              & 16.82(05)    & NTT   \\
13/2/91& 48300.80 & 17.99(05)&  17.74(05)   & 17.16(05)    & NTT   \\
21/2/91& 48308.79 & 19.34(05)&  18.42(05)   & 17.86(05)    & 3.6   \\
24/2/91& 48311.80 & 19.66(05)&  18.55(05)   & 18.05(05)    & D1.54 \\
07/4/91& 48353.75 & 21.32(10)&  20.05(05)   & 19.39(05)    & D1.54 \\
17/4/91& 48363.71 & 21.43(10)&  20.24(05)   & 19.56(05)    & 2.2   \\
20/4/91& 48367.76 &          &  20.26(05)   &              & 3.6   \\
07/8/91& 48475.52 &          &$\le22.50$(20)&$\le22.00(20)$& 3.6   \\
\hline
\end{tabular}

NTT = ESO NTT + EMMI

3.6 = ESO 3.6m telescope + EFOSC1

D1.54 = Danish 1.54m + CCD Camera

2.2 = MPI 2.2m telescope + EFOSC2
\end{flushleft}
\end{table*}

SN~1991D suffers only moderate galactic reddening, E(B-V)$=0.062$
\cite{sch} and there is no evidence of additional extinction in the
parent galaxy. From our best S/N spectra we derive an upper limit of
0.2\AA~ for the interstellar Na~I~D absorptions, which implies
E(B-V)$_{inters} < 0.026$ according to an empirical relation between
these two quantities (Benetti et al 2002a, in preparation). In this paper a
total reddening of E(B-V)=$0.06\pm 0.03$ is assumed.  The parent galaxy
heliocentric velocity is $12 528 \pm 87$ \kms (from NED), which
corresponds to $\mu=36.43(\pm 0.03) - 5log(H_0/65)$.

In Fig. \ref{abs} the absolute V light curve of SN~1991D is compared
with those of the Type~Ib 1999dn and 1990I and Type~Ic SN~1994I. To
match the peak of SN~1991D, the 1994I and 1999dn curves have been
shifted brighter by 2.1 and 3.0 magnitudes, respectively (when the
total reddening is taken into account), while that of SN~1990I
has to be shifted brighter by only 0.2 magnitudes. Figure~3 confirms that
the SN~1991D light curves are as fast as those of SNe~1994I and 1990I.
Moreover, given the V measurements by Remillard et
al. \shortcite{remi} (we have tentatively assigned an
error of 0.3 mag on these measurements) this event reached an absolute
M$_V$ maximum of about $-20.2\pm 0.3$, making SN~1991D the most
luminous SN~Ib yet reported, and among the few reliably measured
supernovae to have been brighter than SNe~Ia (Richardson
et~al. 2002). We want to stress that our own first observation corresponds
to $M_V \simeq -19.0\pm 0.1$; even this is quite bright for a
SN~Ib. Fig. \ref{abs} also indicates that the V light curve possibly reached
maximum at JD$= 48290\pm 2$ (e.g. Feb. 2, 1991).

\subsection{Spectroscopy} \label{spec}

The journal of the spectroscopic observations is given in
Tab.\ref{spec_tab}.  The table lists for each spectrum the date
(col.1), the phase relative to the time of V maximum (col.2), the
instrument used (col.3), the exposure time (col.4), the wavelength
range (col.5), and the resolution as measured from the FWHM of the
night-sky lines (col.6). In order to improve the S/N ratio, in some
cases different exposures, even on consecutive nights, have been
averaged. In such cases the total exposure time is reported.

\begin{table}
\caption{Spectroscopic observations of SN~1991D} \label{spec_tab}
\begin{tabular}{crcrrr}
\hline
\hline
     Date     & phase$^*$ & inst.$^{**}$  &  exp  &   range   
         & res.\\
              & (days)&       & (min) &   (\AA)   &    (\AA)   \\
\hline
12/02/91  &  +10   &  NTT  &  20   & 4050-8050 & 15 \\
21/02/91  &  +19   &  3.6  &  60   & 3800-9700 & 20 \\
20-21/03/91  &  +47   &  1.5  & 135   & 3950-9300 & 15 \\
17/04/91  &  +74   &  2.2  &  30   & 5500-8100 & 15 \\
21/04/91  &  +78   &  3.6  &  60   & 3600-7050 & 20 \\
 \hline
\end{tabular}

* - relative to the estimated epoch of V maximum JD=2448290

** - See note to Table~1 for coding + 1.5 = ESO 1.5m + B\&C
\end{table}

\begin{figure*}
\psfig{figure=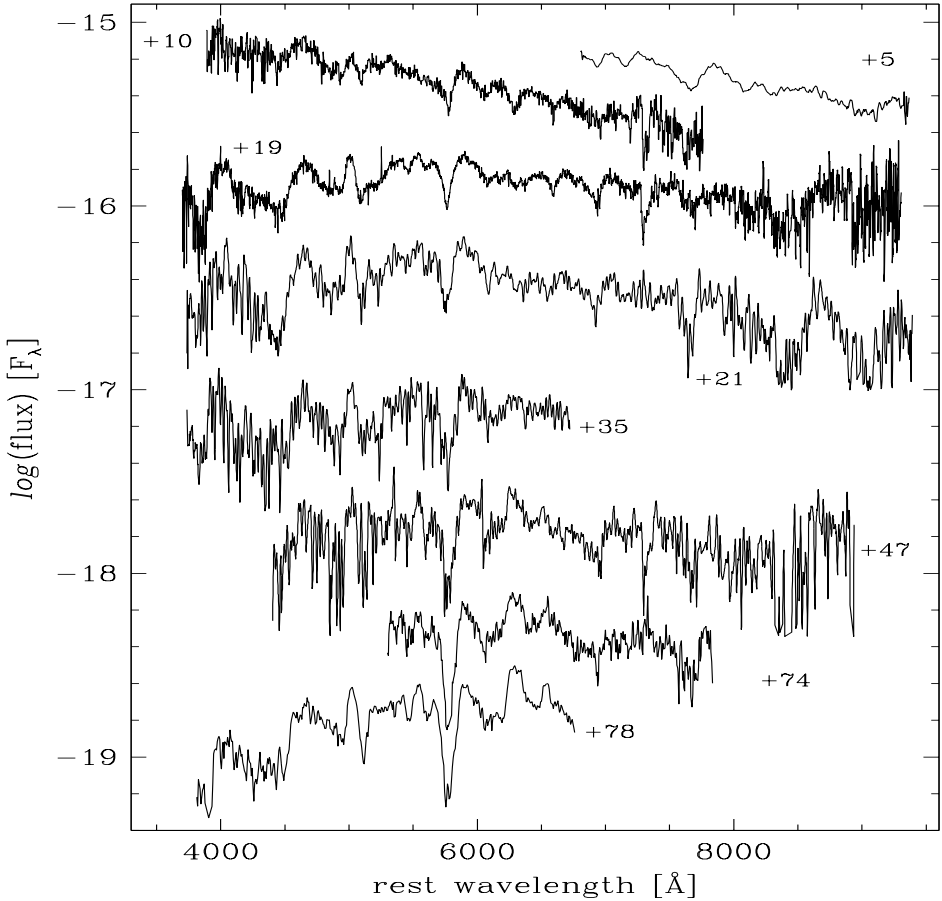,width=13cm,height=13cm,angle=0}
\caption{Spectral evolution of SN~1991D. The +5d, +21d and +35d
spectra are from Matheson et al. (2001). Wavelength is in the parent
galaxy rest frame. The ordinate refers to the +10d spectrum, and the
others have been arbitrarily shifted.}
\label{spec_evol}
\end{figure*}

The flux calibration of the spectra has been checked against the
photometry and in the case of discrepancies the spectra were
adjusted.
Figure \ref{spec_evol} illustrates the spectroscopic evolution of
SN~1991D from phase +5d to +78d.

\begin{figure}
\psfig{figure=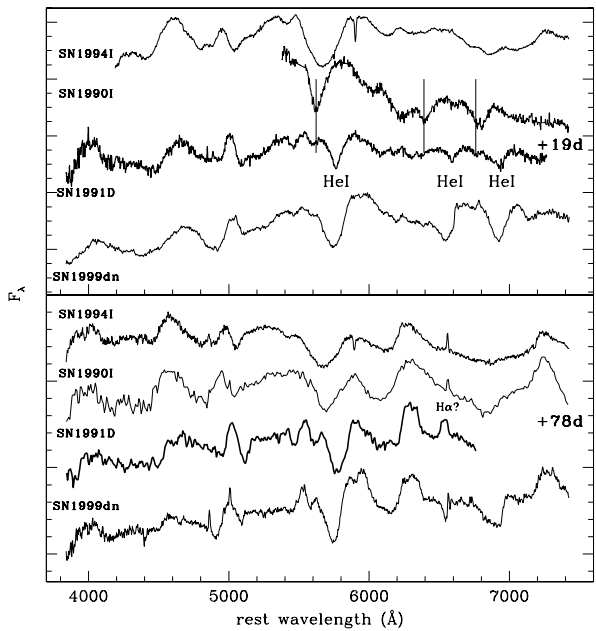,width=9cm,angle=0}
\caption{Top panel: The +19d spectrum of SN~1991D is compared with
spectra of similar phase of the Type~Ic SN~1994I and the Type~Ib
SNe~1990I and 1999dn. Bottom panel: the +78d spectrum is compared with
spectra of the same supernovae at similar phases. The wavelengths are
in the rest frame of the parent galaxies. The main He~I lines have been
labelled. The vertical lines mark the positions of the He~I lines for
an expansion velocity of 13,500\kms.}\label{conf}
\end{figure}

The first two spectra have relatively blue continua with
P--Cygni lines mostly due to Fe~II, Ca~II and He~I.
The presence of He~I $\lambda$5876, $\lambda$6678, and $\lambda$7065 
is clear, but because the He~I lines are weaker in SN~1991D than in 
the events of the B02 sample, and $\lambda4471$ is in a region of 
strong Fe~II absorption, the presence of $\lambda$4471 cannot be 
confirmed. The expansion
velocity as calculated from the deep absorption produced by He~I 5876
and assuming negligible contamination from NaID,
is about 5800 \kms, definitely lower than the velocities shown by
typical SNe~Ib at similar epochs ($\sim 10 000$ \kms, and up to $\sim
13 500$ \kms in SN~1990I).  However, in SN~1991D there may also be
weak highly blueshifted He~I lines with a velocity similar to that
seen in SN~1990I (cf. upper panel of Fig. \ref{conf}).

The spectrum of phase +19d shows a He~I $\lambda 5876$ absorption that
has an unusual triangular shape.  The He~I lines are quite weak
compared to those of a typical SN~Ib such as SN~1999dn (see upper
panel of Fig. \ref{conf}).

The overall appearance of the +78d spectrum is similar to those of
other SNIb/c (see Fig. \ref{conf}) but with some notable
differences. The He~I 5876--NaID absorption, still the dominant feature of
the spectrum and now with a normal absorption shape, shows an
expansion velocity ($\sim 5000$ \kms), lower than that of other
SNe~Ib/c ($\sim 7000$ \kms, cf. Fig. \ref{conf}). The nebular lines
of [O~I] \lam\lam6300, 6364 look normal, with the emission peak at
$\sim 6294$\AA. The [O~I] \lam5577 emission may be present as in
SN~1999dn, the peak occurs at 5544\AA. At about 6537\AA\ there is a
weak but definitely real emission feature with a boxy profile having
FWZI $\sim 98$\AA~ (deconvolved by the spectral resolution), which
could be tentatively identified as \Ha~ with a maximum expansion
velocity of about 4500 \kms.  This emission was already visible in the
noisier +47d and +74d spectra with rest frame peaks at 6523\AA~ and
6543\AA, respectively.  This feature is not normally seen in SNe~Ib/c
(see Fig. \ref{conf}) and could be a sign of circumstellar
interaction. The overall displacement of the main emission features
toward the blue might be an indication of an asymmetry in the
ejecta (dust formation in the ejecta being less probable).

\begin{figure}
\psfig{figure=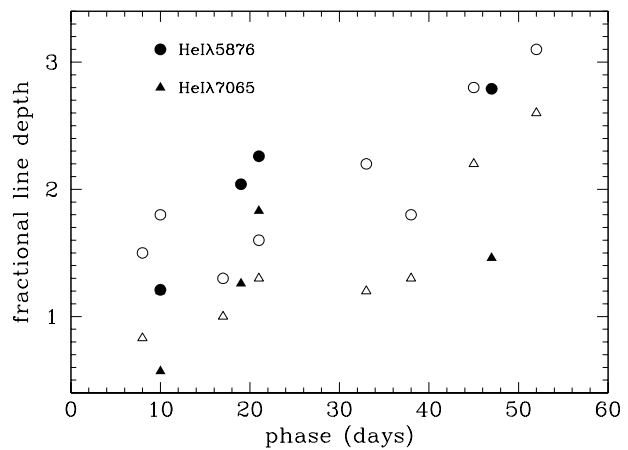,width=9cm,angle=270}
\caption{Temporal evolution of fractional line depths of He~I
$\lambda5876$ and $\lambda7065$ normalized to the fractional line
depth of He~I$\lambda 6678$, as plotted by Matheson et al. 2001. Open
symbols refer to M01 data of SNe~Ib, filled symbols refer
to SN~1991D.}
\label{frac}
\end{figure}

Fig.  \ref{frac} shows that the ratios of the depths of the He~I
absorption features in SN~1991D follow the general trend shown by the
SN~Ib sample presented in M01, even though the SN~1991D He~I lines are
weaker than in normal SNe~Ib.

\section{Synthetic Spectra}\label{syn}

\begin{figure}
\psfig{figure=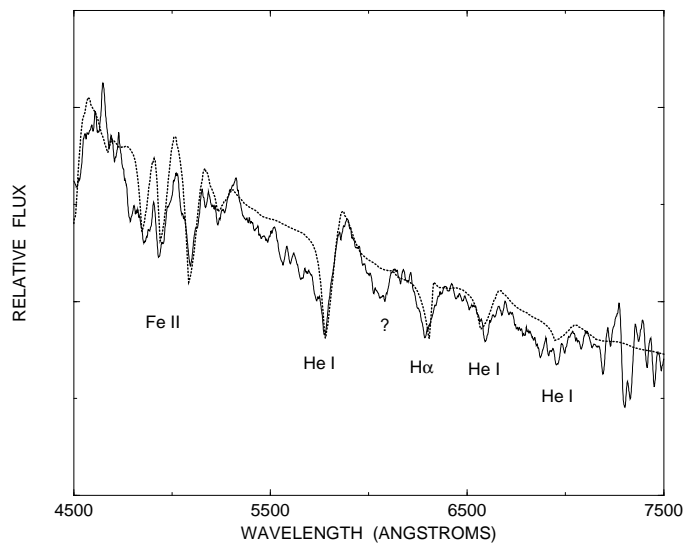,width=9cm,angle=270}
\caption{The +10d spectrum of SN~1991D (solid line) is compared with a
synthetic spectrum (dotted line) that has $v_{phot}=5000$ \kms\ and
$T_{bb}=8500$~K, and contains lines of Ca~II, Fe~II, He~I, and
hydrogen.  The hydrogen lines are detached at 12,000 \kms.}
\label{synh}
\end{figure}

\begin{figure}
\psfig{figure=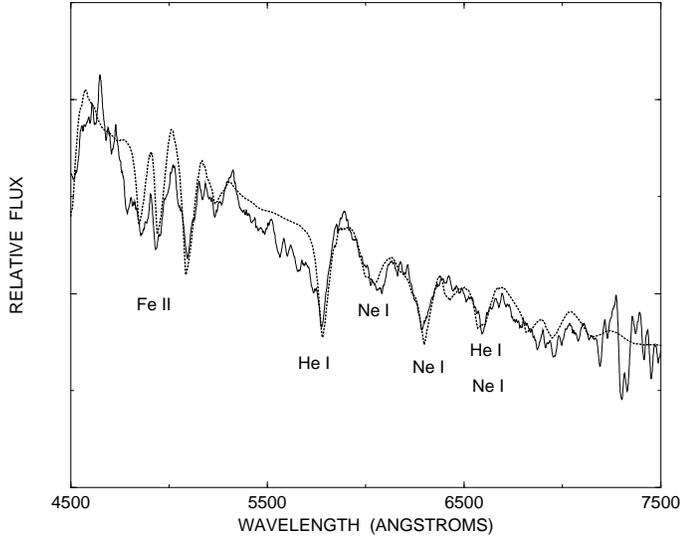,width=9cm,angle=270}
\caption{Like \ref{synh}, except that hydrogen lines are removed and
undetached Ne~I lines are included.}
\label{synne}
\end{figure}

The synthetic spectra shown in Figs. \ref{synh} and \ref{synne} were
computed with the parameterized supernova synthetic--spectrum code
SYNOW.  For a discussion of the physical assumptions and the fitting
parameters see B02.

Fig. \ref{synh} compares the (smoothed) +10d spectrum of SN~1991D with
a synthetic spectrum that has a velocity at the photosphere
$v_{phot}=5000$ \kms\ and a blackbody continuum temperature $T_{bb} =
8500$~K, and contains lines of Ca~II, Fe~II, He~I, all undetached
(i.e., having non--zero optical depths beginning at the photospheric
velocity of 5000 \kms). The value of $v_{phot}$, as determined from
the Fe~II features, is significantly lower than that of the SNe~Ib of
the B02 sample at the same epoch.  For that sample, the characteristic
value of $v_{phot}$ at +10d was 8000 \kms.  Also, in the B02 sample,
at +10d the He~I lines had to be detached; i.e., the optical depths of
the He~I lines were non--zero only above velocities higher than
$v_{phot}$. The detachment velocities ranged from about 8000 to
12,000 \kms at +10d.

In the synthetic spectrum of Fig. \ref{synh}, hydrogen lines also are
included, detached at 12,000 \kms\ so that the H$\alpha$ absorption is
sufficiently blueshifted to match the observed absorption near
6280~\AA\ (the ``6300~\AA\ absorption'' of B02). The mismatch of the
shapes of the observed and synthetic H$\alpha$ features is not
disturbing given our simplifying assumption that the hydrogen line
optical depths rise discontinuously from zero at 12,000 \kms.  The
optical depth of the H$\alpha$ line is sufficiently low that the
identification can not be checked by means of H$\beta$.  Note that the
synthetic spectrum of Fig. \ref{synh} provides no identification for
the observed absorption near 6065~\AA.

Figure \ref{synne} is like Fig. \ref{synh} except that in the
synthetic spectrum the hydrogen lines are removed and undetached lines
of Ne~I are included.  Now the strongest line of Ne~I, \lam6402,
accounts for the 6285~\AA\ absorption, and some other Ne~I lines
account for the observed absorption near 6065~\AA, which was
unaccounted for in Fig. \ref{synh}.  Ne~I lines also slightly improve
the fit to the observed feature near 6590~\AA, which in
Fig. \ref{synne} was accounted for by He~I \lam6678 alone.  Given that
the Ne~I lines are undetached, and therefore do not require a separate
free parameter such as the detachment velocity that is required for
H$\alpha$, the possibility of the presence of Ne~I lines in SN~1991D
must be taken seriously.\\
If Ne~I lines are present they would need to be nonthermally excited, 
as are the He~I lines in SNe~Ib (Lucy 1991; Swartz et~al. 1993).  In 
the synthetic spectrum of Fig.~8 the optical depth at the photosphere 
of Ne I $\lambda$6402 is 2, while Fig.~4 of Hatano et~al. (1999) shows 
that for a helium-rich composition in LTE the optical depth of this 
line peaks at 0.1 at T$\simeq8000$~K and decreases steeply at lower 
temperatures.  Thus the departure coefficient required of nonthermal 
excitation is at least 20, and may be much higher.  Detailed spectrum 
modeling of a well observed SN~1991D--like event would be required to 
determine the departure coefficient. 

In the sample of B02, three SNe~Ib --- SNe 2000H, 1999di, and 1954A ---
showed deep H$\alpha$ absorptions accompanied by weak H$\beta$
absorptions.  For these three events, B02 rejected the Ne~I
alternative to H$\alpha$.  However, the other events of the B02 sample
had weaker ``H$\alpha$'' absorptions that, as in SN~1991D, could not
be checked by means of H$\beta$.  Thus Ne~I remains a possible
alternative identification in these other events.

\section{Results and discussion}\label{res}

SN~1991D shows He~I lines in the spectrum that 
classify it as a Type~Ib. However, it was a peculiar
supernova. Photometrically, it was unusually bright, reaching at least
$M_V = -19.0$ and more likely $M_V=-20.2$.  Spectroscopically, helium
lines were significantly weaker than in typical SNe~Ib (cf. B02) and
the inferred velocity at the photosphere was rather low. In addition
He~I lines had non--zero optical depth down to an unusually low velocity
and some of the spectral features may be attributed to Ne~I.

In order to obtain some information on the progenitor and the
explosion scenario of this peculiar event we used as diagnostic tool a
semi-analytic model originally developed to compute the light curve
and line velocity profile of Type II supernovae (Zampieri et al. 2002,
in preparation). The model has been adapted to Type Ib/c supernovae
including an approximate treatment of the leakage of the gamma-rays
produced by radioactive decay of $^{56}$Co. Despite adopting a number
of simplifying assumptions, such as uniform density, homologous
expansion and constant opacity, comparison with the results of
radiation-hydrodynamic computations has shown that the model is able
to capture the essential physical information and to produce reliable
parameter estimates.

\begin{figure}
\psfig{figure=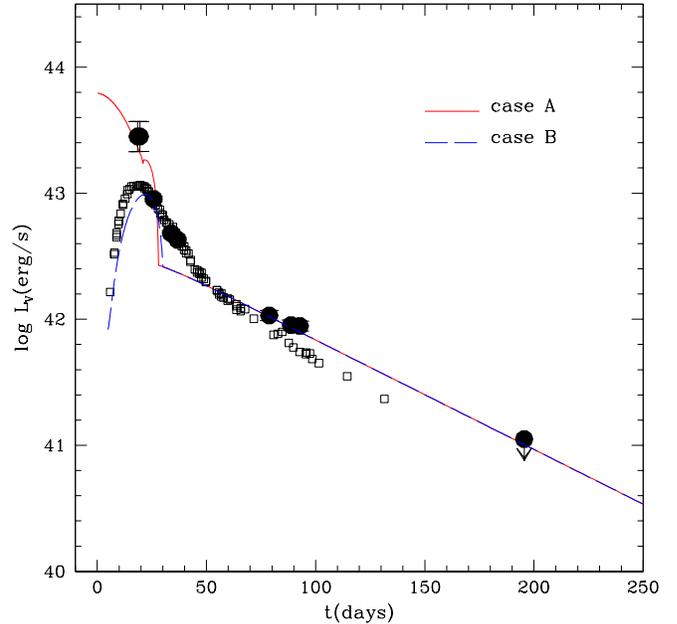,width=9cm,angle=0}
\caption{The absolute V light curve of SN~1991D is compared with our
semi-analytic model (cf. Tab. \ref{th_tab}) and with that of type Ia
SN~1994D (open squares - Patat et al. 1996). The phase is from
explosion that is supposed to occur 15 days before maximum.}
\label{fit}
\end{figure}

The astonishingly high luminosity of SN 1991D during the first $\sim$
20 days ($L_V >10^{43}$ erg s$^{-1}$) and the large luminosity in the
tail ($L_V \simeq 10^{42}$ erg s$^{-1}$ at $\sim 80$ days) are not
typical of Type Ib supernovae and are large even for Type Ia
supernovae. In particular, the first two photometric
points of the discovery the SN impose $L_V \simeq 5\times10^{43}$ erg
s$^{-1}$. This poses serious problems for modeling the light curve
with a C+O or He core of a massive progenitor star that has lost its
outer H envelope. Even if the very high early luminosity may be
associated with emission at or immediately after shock break-out it
still requires a very large initial radius.

\begin{table*}
\caption{Parameters from the semi-analytic model}
\label{th_tab}
\begin{flushleft}
\begin{tabular}{ccccccccc}
\hline
 & radius & ejected mass & $^{56}$Ni mass & $V_0$ & explosion energy & $f_0$ & opacity & $t_{rec,i}$ \\
 & (cm) & ($M_\odot$) & ($M_\odot$) & (cm s$^{-1}$) & ($10^{51}$ erg) &  & (cm$^2$ g$^{-1}$) &  (days) \\
\hline
case A & $10^{14}$         & 1.2 & 0.7 & $10^9$ & 1.4 & 0.5 &         0.1  & 21 \\
case B & $< 10^{12}$     & 2.5 & 0.7 & $10^9$ & 3   & 0.5 &         0.15 &  5 \\
\hline
\end{tabular}

$V_0$ is the velocity of the outermost envelope material

$f_0$ is the fraction of the total initial energy that goes into thermal energy

$t_{rec,i}$ is the time when the bulk of the enevelope starts to recombine

\end{flushleft}
\end{table*}

Because emission from shock break-out is not included in our model, to
bracket possible uncertainties in the parameter estimate, we
calculated semi-analytic models of SN 1991D including (case A), and
neglecting (case B), the first two uncertain photometric points in the
fit. The results of the fit of the light curve and line velocity
profile of SN 1991D are displayed in Table \ref{th_tab} and Figure
\ref{fit}.  The main difference between the two cases is the value
of the initial radius. Fitting the light curve in case A requires a
progenitor with an extremely large initial radius ($\geq 10^{14}$
cm). Yet, in this case the total explosion energy ($E \sim 1.4 \times
10^{51}$ erg) is about one half that required in case B,
essentially because of the lower envelope mass.

In both cases, to account for the high luminosity in the tail, the
explosion that originated the supernova must have been capable of
producing a large amount of radioactive $^{56}$Ni ($\sim 0.7
M_\odot$). This is about 5--10 times larger than the typical yield of
a Type Ib/c supernova \cite{tim}.

We emphasize that the rapid evolution of the early light curve and the
relatively low expansion velocity demand that the ejected envelope
mass be small ($\sim 1-2 M_\odot$). This fact lends support to the
possibility that the progenitor of SN~1991D was a star in a close
binary system (Iwamoto et al. 1994), rather than an isolated very
massive star stripped of the envelope.

We suggest a rather speculative scenario which may explain the main
properties of SN~1991D: a binary system consisting of a white dwarf and a
low mass helium companion ($< 2 M_\odot$). Systems of this type have
been proposed in other contexts to explain the origin and properties
of the emission of cataclysmic variables, supersoft X-rays sources,
and some supernovae (Tutukov \& Yungelson 1996, Iben and Tutukov
1994). Concerning supernovae, two possibilities have been considered:
either the WD accretes matter from the companion and exceeds the
Chandrasekar mass exploding as a type Ia SN (Hachisu et al. 1999); or
the C+O core of the helium star collapses to produce a SN~Ib, if a
substantial fraction of the He envelope is left, or a SN Ic, if it is
lost (Nomoto et al. 1995).

What if the WD ignites while it is hidden inside a He envelope, that
is during a common envelope phase (Tutukov \& Yungelson 1996), or
while it is spiraling into the giant companion (cf. Sparks \&
Stecher 1974)?
Qualitatively, we would expect that because of the large radius of the
helium envelope, the SN would exhibit a high peak luminosity. Yet,
since the envelope mass is small, the luminosity decline would be very
fast. Though camouflaged, this is essentially a type Ia SN and
therefore the mass of radioactive $^{56}$Ni is expected to be large
and the radioactive tail quite luminous. Again, because of the low
ejected mass, gamma-rays from radioactive decays would not be
completely trapped and the late decline of the light curve would also
be fast.
In this respect, we emphasize that the M$_V$ light curve of SN~1991D
is intriguingly similar to that of SN~1994D both in shape and
luminosity, except for the first two brighter points (see
Fig. \ref{fit}).
In these assumptions, the early spectrum is expected to show strong He
lines but, because in the composition structure of a helium star the
second most abundant element after He is Ne (cf. Habets, 1986), the
presence of Ne lines in the spectra would not be surprising. Even the
narrow \Ha~lines observed in the late spectra may be the fingerprint
of the H envelope lost during the first non-conservative mass
transfer/loss episode.

Eventually, when the SN enters the nebular phase, strong [FeII] and
[FeIII] lines should emerge in the spectrum. Unfortunately, the last
spectrum of SN 1991D has been obtained only 78 days after maximum and
is not decisive in this respect.
However we note that spectra of this type have already been observed
in some type Ib/c supernovae. In fact, SN~1993R and SN~1990aj, two
supernovae discovered in the nebular phase, showed peculiar type Ib/c
spectra with very intense iron lines similar to those seen in the
nebular phase of SNIa \cite{macio}.

The fine tuning of the proposed scenario would naturally account for
the rarity of these type of events, which are so luminous that they are
observationally favored in SN searches.

Clearly, to confirm if the proposed scenario is really viable,
detailed numerical radiation-hydrodynamic calculations should be
performed, and detailed non--local--thermodynamic--equilibrium
spectral calculations, including nonthermal ionization (e.g., Baron
et~al. 2000), should be carried out. In particular, these
calculations will show whether Ne I can produce lines of the
appropriate strength.
From the observational point of view, in the fortunate circumstance
that a similar object will be discovered in the future, an effort
should be made to secure spectra in the nebular phase.

The very existence of peculiar SNe like SN~1991D naturally leads to
other issues. Bright SNe of this type might contaminate the sample of
high-z objects used for the determination of the geometry of the
Universe. Although, as speculated above, they might originate from
thermonuclear explosions and share some photometric properties typical
of SNIa, on the other side they would lead to misinterpretation of the
observed data.
In the recent past another bright atypical SN, the peculiar type Ic
SN~1992ar, has been pointed out as a possible contaminator of the SNIa
high-z sample \cite{cloc}. Although the rate of such events in the
local Universe seems relatively small, their frequency at high-z
remains to be determined.

\bigskip

\noindent
{\bf ACKNOWLEDGMENTS}
We acknowledge support from the Italian Ministry for Education,
University and Research (MIUR) through grant Cofin MM02905817.
This work has been also supported in part by NSF grant AST-9986965 and
NASA grant NAG5-3505.
This research has made use of the NASA/IPAC Extragalactic Database
(NED) which is operated by the Jet Propulsion Laboratory, California
Institute of Technology, under contract with the National Aeronautics
and Space Administration.

\end{document}